\documentclass[preprint,12pt]{elsarticle}

\usepackage{graphicx}
\usepackage{adjustbox}
\journal{Neural Networks}

\usepackage[utf8]{inputenc}

\title{On the spatiotemporal behavior\\in biology-mimicking computing systems}

\author
{J\'anos V\'egh\\
	Kalim\'anos BT, Hungary\\
	\'Ad\'am J. Berki\\
	University of Medicine, Pharmacy, Sciences and Technology\\ of Targu Mures, Romania}

\date{18 Sept 2020}

\begin{document}
\begin{frontmatter}
	\begin{abstract}
		The payload performance of conventional computing systems,
		from single processors to supercomputers, reached its limits the nature enables. Both the growing demand to cope with "big data" (based on, or assisted by, artificial intelligence) and 
		the interest in understanding the operation of our brain
		more completely, stimulated the efforts to build biology-mimicking computing systems from inexpensive conventional components and build different ("neuromorphic") computing systems.
		On one side, those systems require an unusually large number of processors, which introduces performance limitations and
		nonlinear scaling. On the other side, the neuronal operation drastically differs from the conventional workloads.
		The conventional computing (including both its mathematical background and
		physical implementation) is based on assuming instant interaction,
		while the biological neuronal systems have a "spatiotemporal" behavior, although conduction time is typically ignored in computational models of neural network function.
		This difference alone makes imitating biological behavior in technical implementation hard. Besides, the recent issues in computing 
		called the attention to that the temporal behavior is a general feature
		of computing systems, too. Some of their effects  
		in both biological and technical systems were already noticed.
		Nevertheless, handling of those issues is incomplete/improper.
		Introducing temporal logic,
		based on the Minkowski transform, gives quantitative insight into
		the operation of both kinds of computing systems, furthermore provides a natural explanation of decades-old empirical phenomena.
		Without considering their temporal behavior correctly,
		neither effective implementation nor a true imitation of biological neural systems are possible.
	\end{abstract}

\end{frontmatter}
	
	\section{Introduction}
	\label{sec:intro}
	The appropriate time handling is vital for the operation 
	of both biological and technical computing systems. 
	Initially, the technical computing was modeled about the biological computing. The technological possibilities, however, limited
	the true imitation of some features, and the extremely fast
	technological development in the electronic industry has strengthened
	the differences between the true and imitated features in a stealthy way.
	The drastically different speed of evolution of those systems
	significantly changed also their relationship.
	
	The tendency is to use an increasing number of conventional processors in technical systems and introduce a new type of workload: artificial intelligence type computations. Both of these tendencies shed light on the importance of\textit{ timed cooperation of their computing elements}. However, neither of those computing systems formulates the task of timing correctly. Technical computing assumes "instant interaction", corresponding to the standpoint of the science a hundred years ago. Its use leads to massive energy-wasting and extremely low payload computing performance when solving real-life computing tasks.
	The biological computing uses a description for the experienced "spatiotemporal" behavior, where the "space" and "time" are handled mathematically as separated functions.
	
	The need for modeling the biological functionality in
	technical computing systems is more and more pressing.
	Imitating their real biological behavior, despite some
	initial successes on small systems, comprising only a few neurons,
	is getting more challenging as the complexity grows. 
	The existence of a limiting interaction speed in both kinds of computing systems,
	enables us to introduce a correct mathematical handling in both fields, and in this way, forms the basis for mutual understanding
	the needs and possibilities of the other field,
	for researchers of both biology and computing.
	
	Given that Heraclitus, BC. 500, stated: "No man ever steps in the same river twice" the idea of the space-time system is ancient.
	In studying biological neural systems, 
	it is known since the beginnings that if we repeat a measurement at a different location in the system, or at a different time at the same place of the system, the measured values are different, although the phenomenon under study is the same. 
	The measurable parameters of an event change their value both in time and space, and we see the same event at a different time in a properly chosen place. This common experience is expressed with the wording, that the systems show a "spatiotemporal" behavior
	\cite{PerturbationNeuralComputation:2002,SpikingNNLiquidSpace:2019}.
	The used phraseology is closely related to the "space-time" coordinates, introduced by Minkowski.
	Despite the evident resemblance, to the authors' best knowledge, no one used the formalism of Minkowski-transform to describe the behavior of biological or technical computing systems.
	The philosophical discussion about "space and time in the brain"~\cite{BuzsakiTheBrain:2019} mentions Minkowski-transform, but we discuss it from a completely different point of view: to describe the information delivery in computing networks, both in biological and technical ones.

	The classic science knows temporal dependence of interactions
	(from our point of view: transferring information via physical interaction)  only in the sense that if we move, for example, an electric charge, the \textit{frequency} of the generated electromagnetic (EM) wave can be calculated. However, because of assuming instant interaction, the \textit{speed} of the EM wave cannot: \textit{the instant interaction is achievable only having infinitely large speed}.
	A similar interpretation was used in former studies in describing neural behavior:
	the mathematical formalism used time ($t$) as an independent parameter that was not connected (through the interaction speed of the action under study) to the spatial coordinates. Because of this, similarly to the case of \textit{frequency} and \textit{speed}, \textit{one of the vital features
		of computing (including neural) networks
		remained out of sight of the research}, resulting in an incomplete description of their behavior.

	The fact that the speed of light is finite was known since Galilei.
	Moreover, Einstein discovered that interactions, such as forces between objects having electric charge or gravitational mass,
	have a finite interaction speed (in other words,
	their interaction is not instant, although very fast). In his (implicit\footnote{
		Einstein in his classic paper "On the electrodynamics of moving bodies"~\cite{EinsteinSpecial:1905}
		speaks about that "light is always propagated in empty space with a definite velocity c," given that the light represents the propagation of the 
		electromagnetic interaction.
		In the abstract of the paper, however,
		he mentions that the speaks about "the phenomena of electrodynamics as well as of mechanics", i.e., gravity. The formalism, however, was available
		for electrodynamics only; for gravity only a decade later.}) interpretation, \textit{there exists a universal limiting speed for interactions, that even the light (given that it is a propagating electromagnetic interaction) cannot exceed}.
	The scientific truth about the existence of a finite interaction speed, in general, was recently confirmed
	by providing experimental evidence for
	the existence of gravitational waves.\footnote{Both assuming the non-instant nature of interactions, and demonstrating their existence, deserved Nobel price.} The experimental evidence also indirectly underpins that \textit{the mathematical background of the Minkowski transform is well-established and correctly describes nature}.  Modern treatments of special relativity base it on the single postulate of Minkowski spacetime~\cite{ModernRelativity:1993}.
	
	In our electronic devices, the EM waves are propagating with speed proportional to light's speed. In the first computer~\cite{EDVACEckertMauchly,GodfreyIEEE1993} (as well as in the present computers), the interaction speed was in the range of $10^8~m/s$-range.
	The "processor size" was in the $m$-range,
	the timings (cycle length, access time, instruction's execution time) in the $msec$-range. Under those conditions, the 
	"spatiotemporal" behavior could not be discovered, and also theoretically, it could be safely neglected. For today, thanks to the technological development, the processor dwarfed million-fold, the timing got million-fold faster. The interaction speed, however, did not change. Since the first personal computers' appearance, the physical components' characteristic size, such as the length of buses connecting their components, did not change significantly (unlike the distance of computing gates, see Moore's observation).
	
	Thanks to the decreasing density and the increasing frequency, \textit{the speed of changing the electronic states in a computing system, more and more approached the limiting speed}.
	It was recognized that the system's clock signals must be delayed~\cite{ClockDistribution:2012}, that effort takes nearly the half of the power consumption of the processor (and the same amount of energy is needed for cooling).
	Furthermore, only about 20\% of the consumed (payload) power is used for computing~\cite{WhyNotExascale:2014}; the rest goes for transferring data from one place to another. Despite these shortcomings, it was not suspected that \textit{the physical implementations of the electronic components have a temporal behavior}~\textbf{\cite{VeghTemporal:2020}}. However, it was the final reason of many  experienced issues, from the payload 
	performance limit of supercomputers~\textbf{\cite{VeghHowMany:2020}} 
	and brain simulation~\textbf{\cite{VeghBrainAmdahl:2019}}, to the weeks-long training time in deep learning~\textbf{\cite{VeghAIperformance:2020,VeghScalingANN:2020}}.
	
	In contrast, in biological computing systems, the "spatiotemporal" behavior of dynamically interacting neurons,
	is explicitly investigated.
	Their interaction speed ("conduction velocity") depends both on their inherent parameters and actual conditions. The $cm$-range distances, the $m/s$-range interaction speed, and the $msec$-range timings (periodicity, spike length, etc.) prove that the proper description of the biological networks is feasible only with temporal logic. In this context, both in theoretical descriptions and real-time\footnote{The real-time in our terminology means that all computing events happen on the biologically correct time scale, instead that, on average, the computing time matches the biological time.} simulations: the temporal behavior is a vital feature of biological systems.
	The "liquid state machine" model grasps the essential point of
	the biological neural networks:  \textit{their logical behavior cannot be adequately described without using
		both time and space coordinates}.
	However, that model handles them separately; it does not connect those coordinates in a way,
	as proposed here. Because of this, that model cannot provide a
	full-featured description of the behavior of biological neural networks, and also it does not enable us to analyze the temporal behavior of their technical implementations.
	
	The Minkowski-transform was famous for its role in accepting, quickly and widely, Einstein's special relativity theory,
	providing an expressive and picturesque mathematical frame for it. The Minkowski-transform, however, is self-contained with assuming only what usually is considered as one of the essential consequences of the special relativity, that a limiting speed exists.\footnote{The paradox consequences of special relativity, such as time dilation and length contraction, are indeed relevant for \textit{observers in different frames}.
		The simplified discussion given here is sufficient for the case of one observer.}

	For our paper, we only assume that a limiting speed (in both electronic and biological systems) exists, and \textit{transferring information in the system needs time}. In biology, the limiting speed (the conduction velocity) is modulated, but our statement holds for any single action: \textit{the spatial and temporal coordinates are connected through the corresponding limiting speed}.
	In our approach, \textit{Minkowski provided a mathematical method to describe information transfer phenomena in a world, where the interaction speed is limited}.
	For example, if we have our touching sense as the only source of information from the external world,
	we need to walk to the object, and this automatically limits the information propagation speed (both touching and being touched) to our walking speed. 
	This case is the same when transferring information in computing systems: the space-time four-coordinates describe that world.
	The only new assumptions we make, are, that the events also have a processing time, such as an atomic transition, executing a machine instruction, or issuing/receiving a neural spike\footnote{Receiving a neural spike, however, is a little bit special case: because of the integration,
		in some cases, the "processing time" can vary.}, furthermore, that the interaction speed is other than the speed of light.
	
	We can proceed, following Minkowski, merely introducing a fourth coordinate,
	and through the assumed limiting speed (without making further assumptions about the value and nature of that speed),
	we can \textit{transform the time of propagation of an event (the interaction, aka the physical transfer of the information) to a distance}, within which the interaction can have an effect in the considered time duration. \textit{Notice the critical aspect, that space and time not only have equal rank, but they are connected through the interaction speed, and that all coordinates have the same dimension}.

	The present paper focuses on three major topics.
	In section~\ref{sec:Minkowski}, it is shown how the (inverse) Minkowski transform can be used to describe both computational and biological neural networks. The section introduces the idea,
	that \textit{both processing time and data transfer time are naturally part of any kind of computing}. 
	
	In section~\ref{sec:Biological} we take the focus narrower, especially to biology, and discuss the importance of time synchronization 
	in collective oscillations, one of the essential biological functionalities of the brain.
	The analysis results in a trivial explanation of the commonly used principle
	"neurons that fire together wire together."
	The present paper does not want to discuss how different biological phenomena (such as pre-synaptic and post-synaptic processing, etc.)
	shall be included in the model.
	
	In understanding the excessively complex operation of the brain, computer methods also have their role. Section~\ref{sec:Technical}  discusses some technical implementations and how their temporal behavior limits the intended faithful imitation of biological neural systems' operation.
	Again, it is out of scope of the paper to analyze special-purpose artificial neuronal chips\footnote{The effects of some inherited solutions, such as "grid time", are discussed in~\cite{VeghScalingANN:2020}} (partly because of the lack of sufficient (proprietary) technical information of their implementation).
	Instead, some common principles are discussed.
	
	The importance of and the need for AI applications speedily grows, and the decades-old truth, that "more is different"~\cite{MoreIsDifferent1972},
	remains out of sight. The section also  
	shortly analyses how closely artificial neural networks
	can mimic biological ones, and what limitations on their performance the 
	methods, inherited from the non-temporal computing, introduce.

	\section{The Minkowski transform\label{sec:Minkowski}}

	As suspected by many experts, the computing paradigm itself, "\textit{the implicit   hardware/software  contract}~\cite{AsanovicParallelCACM:2009}", is responsible for the experienced issues in computing:
	"\textit{No current programming model is able to cope with this development [of processors], though, as they essentially still follow the classical van Neumann model}"~\cite{SoOS:2010}.
	When thinking about "advances beyond 2020", the solution was expected from the "\textit{more efficient implementation of the von Neumann architecture}"~\cite{DeBenedictis_supercomputing:2014}, however.
	
	There are many analogies between science and computing~\textbf{\cite{VeghModernParadigm:2019}}, among others, how they handle time.
	Both classic science and classic computing assume instant (infinitely quick) interaction between its objects.
	That is, an event happening at any location can be instantly seen at all other locations.
	This assumption implies that the information inside a computing system is delivered instantly, and it is always 
	available when the processing unit is ready to process it.
	Mathematics assumes only logical dependence between its operands,
	otherwise assumes that they are instantly available.\footnote{The idea, however, that processing the data does not happen instantly, somehow slipped in computing theory, although it is handled only implicitly.}
	Because of the "contract", engineering implementation
	must follow the same rules of the game; 
	leading (among others) to assuming "weak scaling" that neglects the transfer time even in globe-wide networked distributed systems~\textbf{\cite{VeghScalingANN:2020}.}
	\textit{The time has no specific role in computer science} (in a sense given above).
	In science, discovering that there exists an insurmountable interaction speed led to the modern scientific disciplines' birth. 
	Those new disciplines did not invalidate the corresponding classical ones. Instead, they drew their range of validity and described the world outside that range.
	
	\subsection{Why temporal logic is needed}
	In computing, distances get defined during the fabrication of components and assembling the system. Operating them, however, temporal characteristics are used. In biological systems, nature defines the neuronal distances, and in 'wet' neuro-biology, signal timing, rather than axon length, is the right (measurable) parameter. To describe correctly the temporal operation of a biological or a technical computing system, \textit{we need to find out how much later a component notices that an event (aka a piece of information, such as a spike, a clock signal, or a network package arrival) occurred in the system}.
	
	Computer science is based on logical functions. It assumes that their value does not depend on where and when the functions are evaluated. In other words, it is assumed that all events happen at the same time and in the same place. \textit{It is true for both the timeless world of the mathematics and an infinitely small and infinitely fast, physically implemented computer. However, it is increasingly wrong for physically implemented computers as their physical size increases,
		and/or the data transfer time compared to processing time increases, and/or the change of their electronic states approaches the speed of light.} If we can change
	the interpretation of logical functions to another one, using space-time coordinates\footnote{Formally, we only introduce $bool~function(Minkowski\_coord~x=(0,0,0,0))$, instead of $bool~function(void)$}, we can consider the new technological situation in producing, describing and using computing systems, including biological neural systems
	(importantly: the technical systems imitating them);
	using the solid mathematical background of computer science.
	Furthermore, we can change the design principles correspondingly, to provide more effective systems (and mimicking more closely 
	"biological computing").
	
	We suggest introducing a \textit{temporal logic} (i.e., that the value of a logical expression depends on where and when it is evaluated) into computing. The \textit{reverse} of the Minkowski transform is proposed here: 
	we need to use \textit{a special four-vector, where all coordinates are time values}.
	The first three are the corresponding local coordinates (distance, measured along the path of the signal, from the location of the event, 
	divided by the speed of interaction; plus time contributions such as multiplexing or network hops) having time dimension,
	and the fourth coordinate is the time itself; that is,
	we introduce a \textit{four dimensional time-space} system. The resemblance with the Minkowski-space is obvious,  and the name difference signals the different aspects of utilization.

	\begin{figure*}[t!]
		\includegraphics[width=.8\textwidth]
		{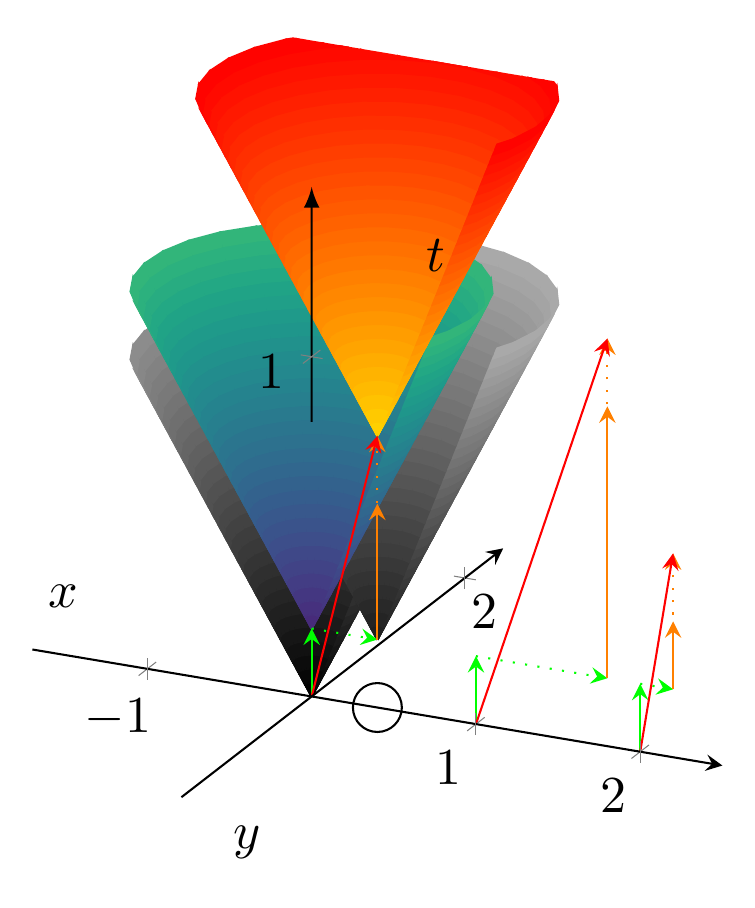}
		\caption{The computing operation in time-space approach. The processing operators can be gates, processors, neurons or networked computers.\label{fig:RelativisticComputation}}
	\end{figure*}
	
	Figure~\ref{fig:RelativisticComputation} shows \textit{why the time must be considered explicitly in all kinds of computing}. The figure shows (for visibility) a three-dimensional coordinate system:
	how an event behaves in a two-dimensional space plus time (the concept is more comfortable to visualize with the number of spatial dimensions reduced from three to two).
	In the figure, the direction 'y' is not used, but enables us to place observers at the same distance from the event, without the need to locate them in the same point.
	The event happens at point (0,0,0); the observers are located on the 'x' axis; the vertical scale corresponds to the time.
	
	\subsection{Reproducing Einstein's hypothetical experiment}
	
	In the classic physical hypothetical experiment, we switch a light on in the origo, and the observer
	switches his light when notices that
	the first light was switched on. 
	If we graph the growing circle (corresponding to the propagation of the light) around the vertical axis of the graph representing time,
	the result is a cone, known as the \textit{future light cone}  (in 2D space plus a time dimension).
	Both light sources have some "processing time",
	that passes between noticing the light (receiving the instruction or a synaptic input)  and switching the light on (performing the instruction or emitting a spike).
	
	The instruction is received at the origo, at the bottom of the green arrow.
	The light goes on at the head of the arrow (at the same location, but later); after the "processing time" $T_p$ passed. Following that, the light propagates in the two spatial dimensions as a circle around the axis "t".
	Observers at a larger distance notice the light at a later time:
	a "transmission time" $T_t$ is needed. The same spike arrives at different neurons (or a signal in a distributed system to different processors) at different times.
	If the "processing time" of the first event's light source were zero, the light would propagate along the gray surface at the origo.
	However, the light will propagate along the blueish cone surface at the green arrow's head because of the finite processing time. 
	
	A circle denotes the position of our observer on the axis "x".
	With zero "transmission time", the second gray conical surface (at the head of the green dotted arrow)
	would describe his light. However, his "processing time" can only begin when the observer notices the light at his position: when the dotted orange arrow hits the blueish surface.
	\textit{A crucial point, that a computing system (including biological
		or artificial neurons) cannot process its data input, until all their data physically arrive at the 
		input ports of the processing unit.} (See the discussion below, how high-speed bus changes the data transmission time and how the fast tensor  processing computes a wrong feedback with uninitialized state variables.) 
	
	At that point begins the "processing time" of the second light source;
	the yellowish conical surface describes the second light propagation.
	The horizontal (green dotted) arrow describes the physical distance of the observer (as a time coordinate),
	the vertical (orange dotted) arrow describes the time delay of the observer light.
	It comprises two components: the $T_t$ transmission time to the observer and its $T_p$ processing time. The light cone of the observer starts at $t=2*T_p+T_t$.
	
	The red arrow represents the resulting \textit{apparent processing time} $T_A$: 
	the longer is the red vector; the slower is the system.
	As the vectors are in the same plane,
	$T_A = \sqrt{T_t^2+(2\cdot T_p+T_t)^2}$, that is  $T_A = T_p\cdot \sqrt{R^2+(2+ R)^2}$.
	That is, \textit{the apparent processing time is a non-linear function 
		of both of its component times} and \textit{their ratio $R$}.
	If more computing elements are involved, $T_t$ 
	denotes the longest transmission time. (Similar statement is valid if the $T_p$ times are different.) Their effect is significant:  if $R=1$, the apparent execution time of performing the two computations is more than three times longer than the processing time.
	
	Two more observers are located on the axis 'x', in the same position.  For visibility, their timings are displayed at points '1' and '2', respectively.  Their results illustrate the influence of the transmission speed (and/or the ratio $R$).
	In their case, \textit{the transmission speed differs by a factor of two}, compared to that, displayed at point '0'; in this way
	three different $R=T_t / T_p$ ratios are displayed.
	Notice that at half transmission speed (the horizontal green arrow, representing the transfer time, is twice as long as that in the origo)
	the vector is considerably longer, while at
	double transmission speed, the decrease of the apparent time
	is much less expressed\footnote{\cite{VeghHowMany:2020} discusses this phenomenon in details.}. 
	
	\subsection{Visualising the temporal operation}
	
	The "liquid state machine" correctly grasps some essential aspects
	of brain operation, but it cannot provide a solid mathematical base for
	describing the details of the time dependence of neuronal operations,
	because \textit{the inhomogeneous coordinates are not connected, they are mathematically separable}.
	In this way, having different conduction velocities, phase-locking or learning
	(and especially those in their technical implementations), remains out of sight of the model.
	Similarly, it is not easy to visualize different relations
	between neuronal state variables.
	\textit{The inverse Minskowski transform works entirely in the temporal domain}
	(as do researchers in 'wet' neurobiology), that is representing
	the neural events in this system is quite natural.
	
	Notice that the handling depicted in Fig.~\ref{fig:RelativisticComputation} \textit{natively separates the
		time-dependent and distance-dependent time components} (and, at the same time, unifies them).
	The processing happens at the same place, and has a time duration, so the corresponding vectors
	are parallel with the time axis. 
	Given that there is no instant interaction,
	a signal, transferring information from one processing unit to another, must change
	both its time and position coordinates, i.e., it can be described
	as a vector which is not parallel with the time axis
	and is not perpendicular to it
	(its projections to the axes are only helping calculation).
	
	\textit{The slope of the vector depends on the interaction speed}.
	The slope can have a proper meaning in technical computing, too, for example in the case of networked distributed systems.
	Nevertheless, in biology, it is a unique feature.
%

	\subsection{The apparent processing time}
	
	Given that the \textit{apparent processing time} $T_A$, rather than the physical processing time $T_p$, defines the performance of the system, $T_p$ and $T_t$ must be concerted.	
	The biology predefines the composition of the time contributions in a biological neuron,
	changing either of the two component times \textit{in their simulation} drastically
	degrades the relevance of simulating their temporal behavior.
	The nature does not solve differential equations, as the computer needs to do, and the technical implementation must convert the "massively parallel" data transfer of biological systems to a sequential transfer. In this way, it can drastically degrade the
	speed (and even their sequence) of transferring information from one processing unit to another.
	That is, both using different mathematical solutions (as well as computing accelerators) and implementing neuronal information transfer changes the ratio between those component times in biology-mimicking systems. As a consequence, \textit{any technical imitation of a biological computing system, has no relation to the real biological operation,
		until it handles the biological time correctly}.
	The apparent processing time $T_A$ and the technical processing time $T_p$ can be shockingly different when using a large number of 
	technical neurons, and transferring the information using 
	conventional technical implementations.
	
	\begin{figure*}
		\includegraphics[width=\textwidth]{
			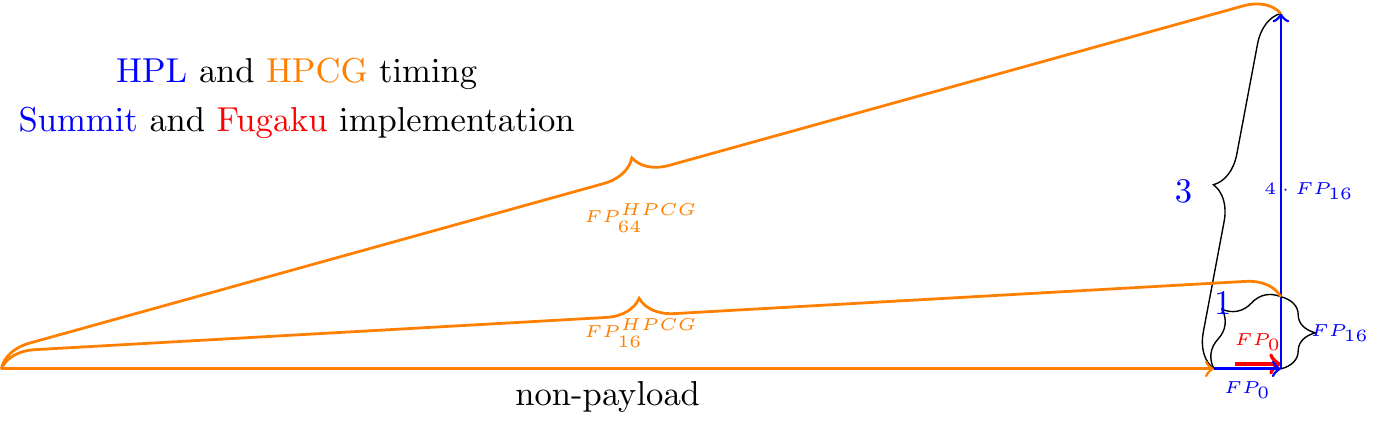}
		\caption{The degrading effect of apparent transfer time $T_A$ on the non-payload contribution, in defining $HPL$ and $HPCG$ efficiencies,
			for double and half precision floating operation modes. For visibility, a hypothetic efficiency ratio $E_{HPL}/E_{HPCG}$=10 assumed. The housekeeping (including transfer time and cache misses)
			dominates, the length of operands has only marginal importance.\label{fig:FP16vsFP64HPCG}}
	\end{figure*}
	
	The mutually blocking effect of computing and transfer operations is demonstrated in Fig~\ref{fig:FP16vsFP64HPCG}.
	Recent supercomputers  $Fugaku$~\cite{DongarraFugakuSystem:2020} and $Summit$~\cite{MixedPrecisionHPL:2018}   provided their HPL performance 
	for both 64-bit and 16-bit operands.
	The power consumption data~\cite{MixedPrecisionHPL:2018} underpin the expectation that their payload performance should be four times higher when using four times shorter operands.
	However, their computing performance shows a moderate performance enhancement because of the needed housekeeping: 3.01 for $Summit$ and 3.42 for $Fugaku$.
	The presence of the (for floating computation)
	non-payload activity acts in the same way as data transmission:
	delays (and in this way: blocks) floating computation. 
	Given that the inverse Minkowski transform
	enables us to handle them uniformly,
	the time delays due to data transmission
	(including physical delivery and data access time; in the spatial domain) and the time delay due to addressing, incrementing, branching 
	(happening at the same place, in the time domain),
	combines in this simple way.
	The slight difference in the contribution of housekeeping 
	(denoted by $FP_0$ in the figure), however, results in a payload performance-limiting factor about three between the two supercomputers.
	In terms of temporal operation, the higher $FP_0$ contribution results in a
	higher apparent processing time, i.e., in lower performance,
	as discussed in detail in~\textbf{\cite{VeghHowMany:2020}}.
	
	The temporal behavior also explains, why different workloads
	(represented by the benchmarks HPCG and HPL, respectively)
	result in different payload performances: the "sparse" operation of the
	HPCG algorithm, furthermore its need for a more intense communication. Increasing the apparent transfer time degrades the system's payload performance.
	In the figure, for visibility, a hypothetic efficiency ratio ten is assumed. The real (measured) efficiency ratio is about 100-250: the cache failures enormously increase the average data transfer time, which blocks the computation, and in this way, increases the apparent processing time.  As estimated in~\textbf{\cite{VeghHowMany:2020}}, 
	in the case of imitating neuromorphic operation on a conventional architecture, this ratio is expected to be above 1000.
	It was bitterly admitted, that "\textit{artificial intelligence, \dots it's the most disruptive workload from an I/O pattern perspective.}"\footnote{ https://www.nextplatform.com/2019/10/30/cray-revamps-clusterstor-for-the-exascale-era/} Notice also, that \textit{under AI workload,
		using half-precision instead of double precision, reduces the power consumption of the system, but 
		has only marginal effect on its computing performance}, see Fig.~\ref{fig:FP16vsFP64HPCG}.
	
	\subsection{Concerting the component times}

\begin{figure*}[htp]
		\maxsizebox{1.0\textwidth}{!}{
			\begin{tabular}{c}
				\hspace{-1.cm}
				\includegraphics[scale=1.5]{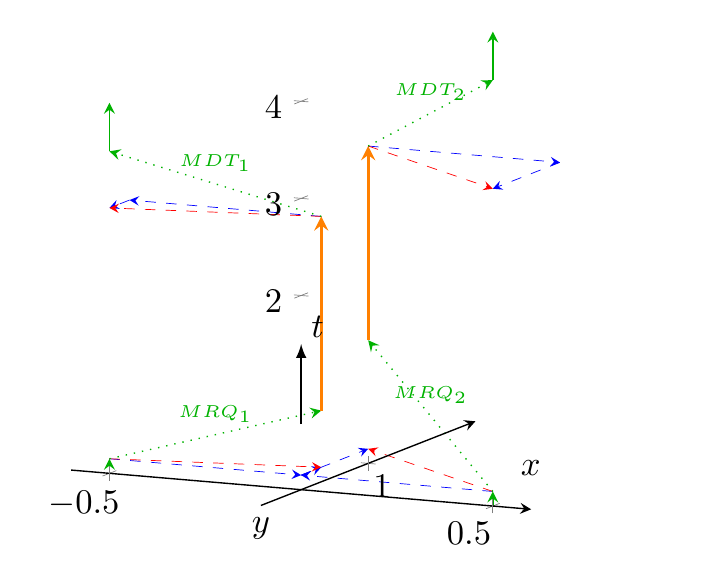}
				~
				~\hspace{-3.5cm}
				\includegraphics[scale=1.5]{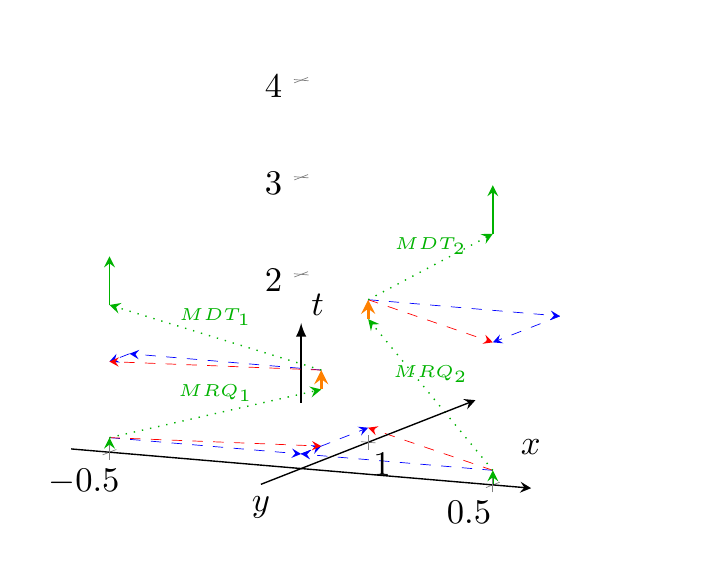}
			\end{tabular}
		}	
		\caption{Performance dependence of an on-chip cache memory, at different cache operating times, in the same topology.
			Cores at (-0.5,0) and (0.5,0) positions access on-chip cache memories at (0,0.5) and (0,1), respectively. Vertical orange arrows represent physical cache operating time. Cache memories, from left to right,
			have physical access speed (on some arbitrary scale) 1 and 10, respectively. Vertical green arrows (from the bottom of the lower arrow to the top of the upper arrow) represent the apparent access time.\label{fig:CachePerformance}}    
	\end{figure*}

	Given that its \textit{apparent processing time} $T_A$ defines the payload performance of the system, $T_p$ and $T_t$ must be properly concerted.
	Fig.~\ref{fig:CachePerformance} demonstrates how the apparent access time
	$T_A$ of an on-chip cache memory changes, if one changes the processing speed $T_p$ of the cache.
	In the figure, two different topologies and two different 
	physical cache operating speeds are used. 
	Two cores are in positions (-0.5,0,0) and (0.5,0,0), furthermore two cache memories at (0,0.5,0) and (0,1,0). 
	The signal, requesting to access the cache, propagates along the dotted green vector
	(it changes both its time and position coordinates; recall that position coordinates are also mapped to time), the cache starts to operate only when the green dotted arrow hits its position. 
	Till that time, the cache is idle waiting.
	After its operating time (the vertical orange arrow), the result is delivered back to the requesting core.
	This time can also be projected back to the "position axes", and their sum (thin red arrow) can be calculated.
	Similarly, the requesting core is also "idle waiting" until the requested content arrives.
	
	The physical delivery of the fetched value begins at the bottom of the lower thick green arrows, includes waiting (dashed thin green lines), and finishes at the head of the upper thick green vector; their distance defines the \textit{apparent cache access time} that, of course, is inversely proportional with the \textit{apparent cache access speed}. 
	Notice that the \textit{apparent processing time} is a monotonic function of the
	physical processing speed, but because of the included 'transmission times'
	due to the physical distance of the respective elements, their dependence is far from being linear.
	The \textit{apparent cache speed} increases either if the cache is physically closer to the requesting core or if the cache access time is shorter (or both).
	The apparent processing time (represented by vertical green arrows) is only slightly affected by the physical speed of the cache memory (represented by vertical orange arrows).
	The concrete example explains why the clever placing of its cache memories resulted in that supercomputer $Fugaku$ outperformed its predecessor by a factor of three.
	
	The case is similar to other components in the computing chain, too.
	In addition to using shorter operands, another plausible assumption is that if we use quick analog signal processing
	to replace the slow digital calculation, as proposed in~\cite{RecipeMemristor:2020,NatureBuildingBrain:2020},
	the system gets proportionally quicker.
	As the temporal behavior explains, a system comprising memristors will be only marginally quicker, until the time of data transfer can be significantly reduced.
	The case is analog with the one shown in Fig.~\ref{fig:FP16vsFP64HPCG}. In the case of a neuromorphic workload, the data transfer $T_t$ dominates the system's payload performance, so even decreasing $T_p$ to zero could not lead to a significant performance enhancement. 
	
	Also, adding analog components to a digital processor has its price. Given that the digital processor cannot handle resources outside of its world, one must call the OS for help. That help, however, is rather expensive in terms of execution time. The required context switching
	takes time in the order of executing $10^4$ instructions~\cite{armContextSwitching:2007,Tsafrir:2007}, which greatly increases
	the total execution time and makes the non-payload to payload ratio much worse.
	In terms of temporal analysis, the ratio of the component times changes to a value range of $R\approx 10^4$. In this way, the frequent context switches will dominate the payload performance of the system.
	
	These cases seem to be very different. They, however, share at least the common feature, that they change not only one parameter: \textit{they also change the
		non-payload to payload ratio} that defines the systems' payload efficiency.
	
	\section{Manifestations of the role of time in biology\label{sec:Biological}}
	The brain maintains different temporal characteristics of its neural events~\cite{PrinciplesNeuralScience:2013}. 
	Moreover, we want to emphasize why adopting their timely behavior is necessary for authentic/successful technical imitations. 	
	The biology-mimicking computing principles seem to be cleared up to a level that even some computing primitives could be proposed~\cite{BrainInspiredBlocks:2020}. When considering the role of time in (all kinds of) biology-mimicking systems,
	some basic principles (and, consequently, those primitives) must be pinpointed or extended.
	As an example, the periodic oscillation (or more precisely: the synchronized co-oscillation) is discussed, which -- in addition to its role in the operation of the brain-- has other roles in biology~\cite{WilliamsPhaseCoupling:1992}. 
	 The phenomena seem to be well-studied also by mathematical methods, see~\cite{CoherentSpatioTemporal:2018} and its cited references.

	Given that the single neurons are embedded in oscillatory networks,
	they can (and shall) synchronize their temporal operation using their low-precision "local clock". Receiving a base frequency,  resets the "local clock" frequently, and the \textit{relative time} can be accurate for a more extended operating period.
	Notice, that neurons at different Minkowski-distance "see" the local frequency
	at different Minkowski-times, so they set their internal phase angle to the
	Minkowski-distance. That is, "at the same time" means different absolute times\footnote{ Concerning Einstein's hypothetic experiment: the neuronal interactions have a speed in the range up to $10^{2}~m/s$;
		this speed is the speed limit of receiving information by the "internal observers" (the neurons) \textit{only}. The scientist, however, is "external observer", having information transfer speed in the range $10^{8}~m/s$} for different neurons, depending on their location: the biology uses Minkowski four-coordinates.
	
	The brain oscillates continuously at a vast range of frequencies (0.02-600 Hz)~\cite{BuzsakiRhythms:2006}, maintaining its neurons' global function at various time scales. The frequency bands are generated by the cell assemblies' current behavior, representing their involvement in different computational processes.
	The phase has a unique role in stationary oscillations,
	as a kind of \textit{fifth coordinate} in space plus time system,
	given that the time duration/phase is not included in the separately handled space+time coordinate system. 
	"\textit{The phases of their oscillations at different frequencies is also relevant in keeping the operation of neurons synchronous.
	A conduction delay of 5ms could change interactions of two coupled oscillators at the upper end of the gamma frequency range
	($\approx 100~Hz$) from constructive to destructive interference; delays smaller than 1~ms could change the phase by $30^\circ$,
	significantly affecting signal amplitude"~\cite{RoleOfMyelinPlasticity:2014}}.

	The initially different phases are tuned to produce a 
	sufficiently strong signal due to the proper superposition of the signals: the Minkowski-time of the signals sent by the peer neurons is adjusted so that the signals arrive at the same Minkowski-coordinate.
	The goal is to produce a maximum current signal amplitude (or integrated charge) at the target 
	(at given time-space coordinate!). Given that the contributing neurons have different time-space coordinates, their phases (the time coordinate) indeed are not necessarily the same. The spatial Minkowski-coordinates are defined anatomically.
	However, as they are located in a way to minimize the needed energy efficiently, their Minkowski coordinates differ usually only marginally.
	Notice that when the phases of the synaptic inputs change, they affect \textit{when} the neuron in question fires (due to weighting,
	different amount of charge is integrated with respect to time, and so the threshold potential is exceeded at a different time).

	The spikes' conduction speed is determined by the axon's anatomical properties (diameter and the amount of myelin) and the type of the synapses (chemical or electrical).    
	The neuronal communication means the transfer of information between distinct nodes. Although the neurons' impulses to muscles could travel with 100 m/s, the interaction speed between two neurons can be found mainly in the $1~m/s$ range. Given that the brain's size is in the $10~cm$ range, the typical neural action speed is placed in the several $ msec $ range, resulting in up to $10~Hz$ maximum operating frequency, without synchronization. 	Hence, it is a challenge to deliver a spike from a distant neuron to its target that receives spikes also from nearby peer neurons (a local assembly) at the right time.
	In more complex tasks, requiring even more extensive areas to cooperate, the problem to be solved becomes even more challenging.

	\subsection{Synchronization by oscillations}
	
	The coordinated operation of neurons is brought about by synchrony sustained by functionally interconnected networks' oscillatory behavior. To gain efficacy, synchrony is established in brief temporal windows. However, their duration depends on membrane resistance, conductance, and other factors. In this periodically emerging time window, the incoming stimuli' response is also influenced by the previously integrated inputs. 
	
	The brain uses temporal packages brought about by oscillatory cycles to overcome these difficulties, i.e., the information is transferred within well-defined time-windows. To support this, 
	the brain developed a unique synchronization method to enable the individual neurons (having different Minkowski coordinates)
	to cooperate in such a way. 
	Given that the goal of their cooperation is, that their collective effort
	can be synchronized on a remote set of neurons;
	interdependent, but distinct, cell assemblies are wired together by long-range neurons endowed with unique properties (thicker myelin sheaths), enabling the establishment of 0-lag synchrony between neuronal populations.

	The first part of the task of synchronization is solved by issuing  some fundamental frequencies
	(aka oscillations) to which the neurons can synchronize themselves. 
	To transfer those "time base" frequency signals with the typical speed of the system would be too slow: it would not enable to distinguish whether a given signal value from a signal having a higher frequency (sampled at a given Minkowski coordinate) belongs to which period of the sampled signal.
	In biological systems, 
	instead of the ionic current conduction, a pure electric conduction technology was developed: the Ranvier cells polarizing each other's input with their output membrane (actually, the neurons have the corresponding 'half' Ranvier cell in the sender and receiver neuron, respectively). However, this "charge cloning" works only if the connection line is not leaking, so these axons are thickly myelinated. The result is that while the potential induced by polarization enables one to see an effect that "\textit{the action potential jumps from one node to another}" (generating local currents only), its speed is significantly (up to about 60 times) higher
	than that of the ionic conduction. 
	In other words, \textit{the biology modulates the interaction speed that permits signals from a distant neuron  (at a more considerable space-only distance) to reach their targets in time, despite that the signals they control arrive from neurons at smaller space-only distance}.

	The receiving neuron could integrate this current in the same way as the input from the others: after the last "half Ranvier cell", the sender (and the path the signal traveled) is no longer known. However, it would be only one of the many (maybe thousands) synaptic inputs, and because of the "voting," it would not have the ability to provide the needed "time base" signal.
	To reach the  required goal, biology developed a method that considers the "rate of rise" of the
	input synaptic current~\cite{LosonczyIntegrative:2006}, too. \textit{Exceeding that  current threshold,}
	\textit{triggers spiking, independently from the state of the other synapses and the action potential.} After spiking, the neuron is "reset", that is,
	its phase is set to zero value.
	That means, that in addition to the \textit{voltage threshold
		for action potential} of  the neuron, one more \textit{current threshold for synaptic currents exists}.
	
	The first part of the discussion in this section
	provides a native explanation for the decades-old empirical rule:
	"neurons that fire together wire together."
	The oscillators must have the same time constant ($RC$ value), to oscillate together. Using the self-frequency of the oscillators
	requires the minimum energy to operate their set; this is required
	to operate the brain with its outstanding efficiency.
	Given that the structure of the members of the assembly are anatomically identical, it is  a plausible assumption, that
	they have identical $RC$ value.
	
	To achieve maximum effect on a remote neuron's synapses, 
	the spikes emitted by a neuronal assembly members shall arrive at the same $phase$ to the target neuron.
	In the assembly, the sender neurons can be "reset" by a "central clock", a base frequency of the brain. That is, the axons have a common source and a destination that branches to the individual neurons only at the very end of their path. Till that point, they usually "wire together."
	However, their Minkowski-distances are slightly different, both from the source to one of the member neurons in the assembly, and from that neuron to the target neuron. The arrival of the synchronizing pulse (a base synchron spike, delivered using the "saltatoric" way of charge delivery), through its path,
	delivers the correct Minkowski-distance for resetting the member neuron, and the neuron has no way (and no reason) to change it.
	
	The Minkowski-distance for the second portion of the triggering can also be (slightly) different. To adjust their \textit{phase} properly, to achieve maximum contribution at the target, the participating member neurons must \textit{learn} how to adjust it.
	As discussed, the same amount of synaptic current contributes differently to the action potential in the member neurons, in function of the phase compared to that of the inputs from other neurons, so \textit{changing the weights (due to the feedback the member neurons receive) leads to a change in the discharge time in the spiking neuron}.
	That is, the feedback is based on the \textit{phase} at which the spike arrives at its destination. After receiving the feedbacks, the change is made in the \textit{time of firing} at the sender. 
	Again, achieve a minimum in the consumed energy (a maximum of the superimposed charges) of this collective operation, at the target, the post-synaptic spikes must arrive at the same local "phase" (concerning the spike produced by the target neuron).
	Given that their Minkowski-coordinates differ only marginally, 
	it looks like that the member neurons spike simultaneously, i.e., they "fire together."

	\begin{figure}[htp]
		\centering{
			\includegraphics[scale=1]{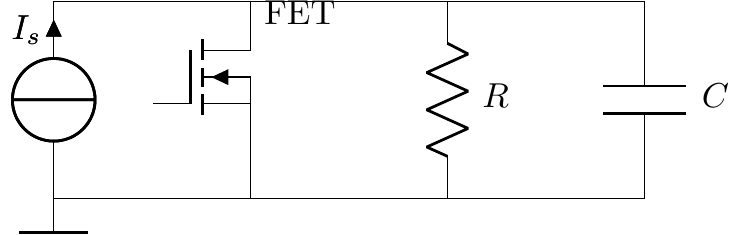}
			\caption{A suggested modified RC circuit
				to consider saltatoric conduction\label{fig:FET}}
		}
	\end{figure}

	The case can be best understood from Figure~\ref{fig:FET}, where a possible technical implementation of the saltatoric behavior is shown (although in this way, only one synaptic input is modeled).
	A field-effect transistor is added to the usual resistor and capacitor that define the frequency of the oscillation in the circuit. As long as the synaptic input current does not exceed the critical value,
	there is no change in the simulated operation. When the input (synaptic) current, however, generates a voltage on the resistor, that opens the transistor,
	the condensator 
	quickly discharges (through a very low limiting resistance), and a spike is produced. (In the biology,
	a refractory period follows, when the input charge and current is grounded; 
	the technical neuron needs no relaxation so that the operating cycle 
	can start immediately.) Anyhow, such a discharge synchronizes the
	"local clock" to the central time base, independently from its "local phase", and makes the operation of the neuron phase-locked to the time-base frequency (RC defines the frequency, although with low precision, and the saltatoric reset defines its phase, precisely).
	Worth to notice that the network of neurons 
	automatically uses the correct Minkowski-time:
	neurons at different places do not necessarily have the same 
	Minkowski-time value, although their functional grouping may cover this difference.

	\subsection{Time and learning}
	From a simplified modeling point of view, learning means to change the weights of the input (synaptic) signals to optimize the operation for the given goal. Changing the weights, however, results in changes in the time of their spiking. Given that the members of an assembly are anatomically identical, in this context only the phase is significant:
	the efficiency requires that the spikes arrive at their destination at the same phase (in this context, it means "at the same time").
	\textit{In models, not considering the timing of spiking,
		the resulting 'phase' feedback is attributed to other input signals, and the effect is distributed among the other (non-temporal) parameters}. Changing the weights of neurons' signals, on one side, means that the learning method attributes improper weights to all of its signals, while it does not consider their timing at all. On the other side, it also means that a portion of the generated synaptic input charge shall be grounded, when only the weights are changed.
	If the learning method considers timing,
	the same operation of the neurons could be reached using less generated charge, i.e., using less energy.
	
	The biology can modulate conduction speed, at the price of
	making structural changes: the corresponding axons must be heavily myelinated.
	One side, making such a structural change, is a prolonged process compared to synchronization. On the other side, when using this latter method of changing the timely behavior of neurons frequently, \textit{it may be less expensive (in terms of energy consumption),
		to accelerate (the produced) less charge than to produce and deliver much more ions}.
	This is why nature combines these two methods: 
	the thickness of the myelin layer can be different, and leads to different conduction speeds. The set of neurons, used to solve a new task,
	can be configured quickly via earthing (i.e., wasting) part of the produced charge, but when using this new experience in a longer time duration,
	it may be optimal to modify the structure via myelination, and in this way to reduce the need of energy for producing and transmitting ions in the long run.
	Also "\textit{axon-specific adjustment of node of Ranvier length is potentially an energy-efficient and rapid mechanism for tuning the arrival time of information in the CNS}"~\cite{RanvierLength:2017}.
	 A model that is worth checking when studying short and long-term learning.
	 	The self-reconfiguration ability of the brain
	 is underpinned by mathematical modeling~\cite{NeuralSelfReconfiguration:2017}, too.

	 Anyhow, the arrival time of information plays a major role,
	 but the technical neurons, representing neuronal states
	with voltage levels, miss all kinds of temporal information.
	Without handling temporal information, vital for operating the biological systems correctly, it is impossible to produce accurate biology-mimicking simulations. Even the timestamps delivered in the spikes can not include the phase: integrating in a time slot is not sensitive to the phase. In technical systems
	(including many-thread simulations on supercomputers),
	the "spatiotemporal" behavior is not considered at all.
	In biology-mimicking systems (including the "liquid state machine") the "spatiotemporal behavior"  is handled in a way where time and place are separable. That is, they are not connected in a way as proposed here.
	
	\section{Technical implementations\label{sec:Technical}}

	\begin{figure*}
		\includegraphics[width=.95\textwidth]{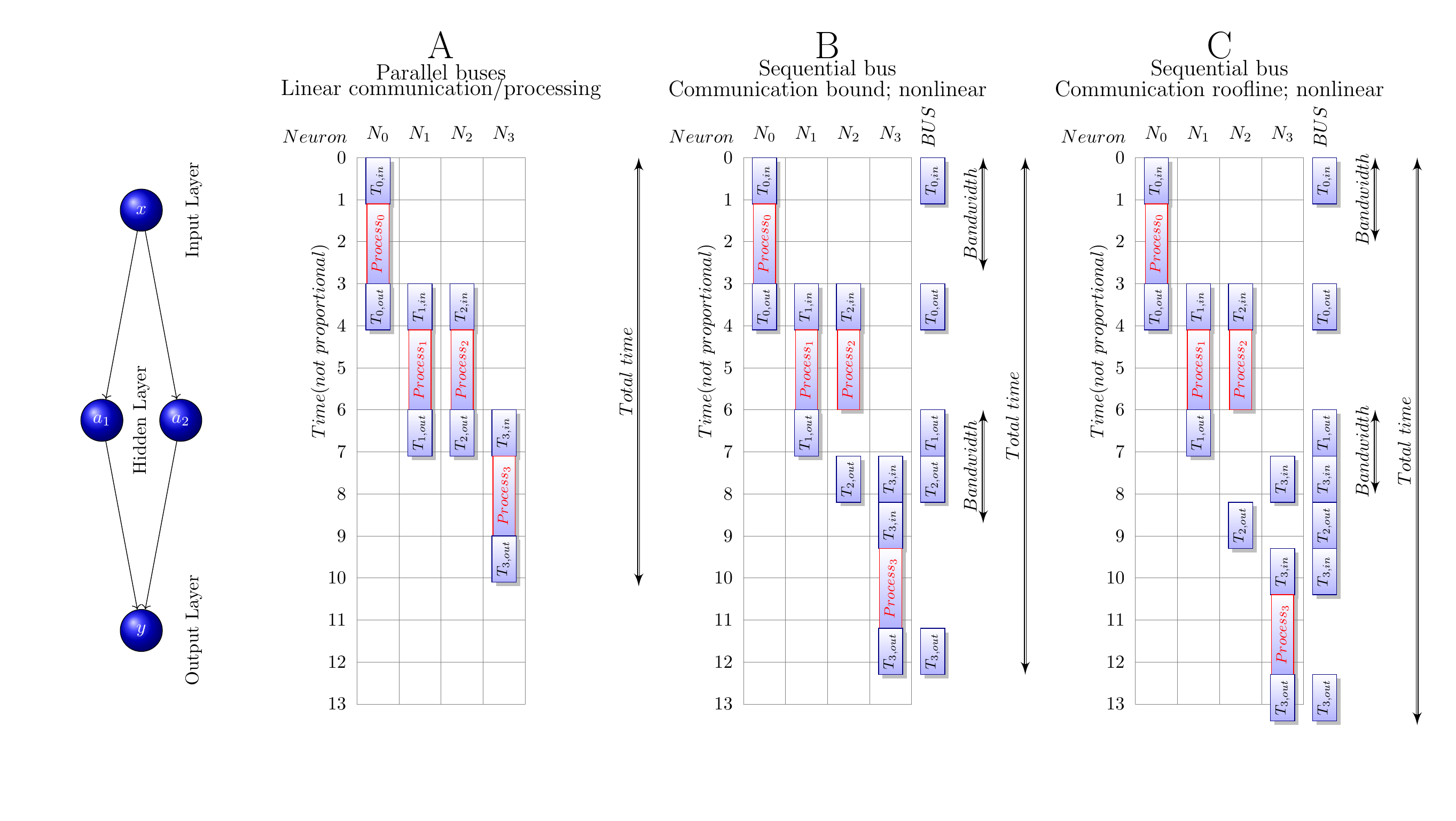}
		
		\vspace{-2\baselineskip}
		
		\caption{Implementing neuronal communication in different technical approaches. A (the biological implementation): the parallel bus; B and C(the technical implementation): the shared serial bus, before and after reaching the communication "roofline"~\cite{WilliamsRoofline:2009}.\label{fig:Neuronal}
		}
		
		\vspace{-\baselineskip}
	\end{figure*}
	
	\subsection{Data delivery method}
	The components of technical computing systems
	(including biology-mimicking neuromorphic ones) are connected through a set of wires, called "bus".    
	The bus is essentially the physical appearance of the "technical implementation" of communication, stemming from the Single Processor Approach~\cite{AmdahlSingleProcessor67}, as illustrated in Fig.~\ref{fig:Neuronal}. 
	The inset shows a simple neuromorphic use case: one input neuron and one output neuron communicate through a hidden layer, comprising only two neurons.
	Fig.~\ref{fig:Neuronal}.A mostly shows \textit{the biological implementation: all neurons are directly wired to their partners}, i.e.,
	a system of "parallel buses" (axons) exists. Notice that the operating time also comprises two "non-payload" times ($T_t$): data input and data output, which coincide with the non-payload time of the other communication party. The diagram displays the logical and temporal dependencies of the neuronal functionality.
	The payload operation ("the computing") can only start after its data is delivered (by the, from this point of view,
	non-payload functionality: input-side communication), and the output communication can only begin when the computing finished. Importantly, \textit{communication and calculation mutually block each other}. 
	Two important points that neuromorphic systems must mimic noticed immediately: i/ \textit{the communication time is an integral part of the total execution time}, and ii/ \textit{the ability to communicate is a native functionality} of the system.
	In such a parallel implementation, \textit{the performance of the system}, measured as the resulting total time (processing + transmitting), \textit{scales linearly with increasing either the non-payload communication speed or the payload processing speed}.

	Fig.~\ref{fig:Neuronal}.B shows a \textit{technical implementation of  a high-speed shared bus} for communication.
	To the right of the grid, the
	activity that loads the bus at the given time is shown.
	A double arrow illustrates the communication bandwidth, the length of which is proportional to the number of 
	packages the bus can deliver in a given time unit.
	We assume that the input neuron can send its information in a single message to the hidden layer; furthermore, the processing by neurons in the hidden layer both starts and ends simultaneously. However, the neurons must compete for accessing the bus, and only one of them can send its message immediately, the other(s)
	must wait until the bus gets released.
	The output neuron can only receive the message when the first neuron completed its sending.    
	Furthermore, the output neuron must first acquire the second message from the bus, and the processing can only begin after having both input arguments. 
	\textit{This constraint results in sequential bus delays both during non-payload processing in the hidden layer and payload processing in the output neuron}.
	Adding one more neuron to the layer introduces one more delay.
	
	At this point, two wrong solutions can be taken: either the second neuron
	must wait until the second input arrives (in biology, a spike also carries a synchronization signal, and triggers its integration), or (in "technical neurons", using
	continuous levels rather than pulses, 
	this synchronization facility is omitted) changes its output continuously,
	as the inputs arrive, and its processing speed enables. 
	In the latter case, however, \textit{until the second input arrives (and gets processed)
		the neuron provides an output signal, differing from the one
		expected based on the mathematical dependence}. As discussed in detail
	in~\textbf{\cite{VeghTemporal:2020}}, this, temporarily may be wrong,
	output signal is known in the electronics, and those "glitches" are eliminated via using a "worst-case" delay for the output signal.
	However, including a serial bus in that computation
	would enormously prolong the needed "worst-case" delay.
	
	Using the formalism introduced above, $T_t=2\cdot T_B + T_d + X$, i.e., 
	the bus must be reached in time $T_B$ (not only the operand delivered to the bus, but also waiting for arbitration: the right to use the shared bus), twice, plus the physical delivery  $T_d$ through the bus.
	The $X$ denotes "foreign contribution": if the bus is not dedicated for "neurons in this layer only", any other traffic also loads the bus: both messages from different layers and the general system messages may make processing slower (and add their contribution to faking the imitated biological effect).
	
	Even if only one single neuron exists in the hidden layer, it must use the mechanisms of sharing the bus, case by case. The physical delivery to the bus takes more time than a transfer to a neighboring neuron (both the arbiter and the bus are in $cm$ distance range, meaning several $nsec$ transfer times, while the direct transfer between
	the connected gates may be in the $psec$ range).
	If we have more neurons (such as a hidden layer) on the bus and work in parallel, they must all wait for the bus. 
	The high-speed bus is very slightly loaded when only a couple of neurons are present. Its load increases linearly with the number of neurons in the hidden layer (or, maybe, all neurons in the system).
	The temporal behavior of the bus, however, is different.

	\begin{figure}
		\includegraphics[width=.8\columnwidth]
		{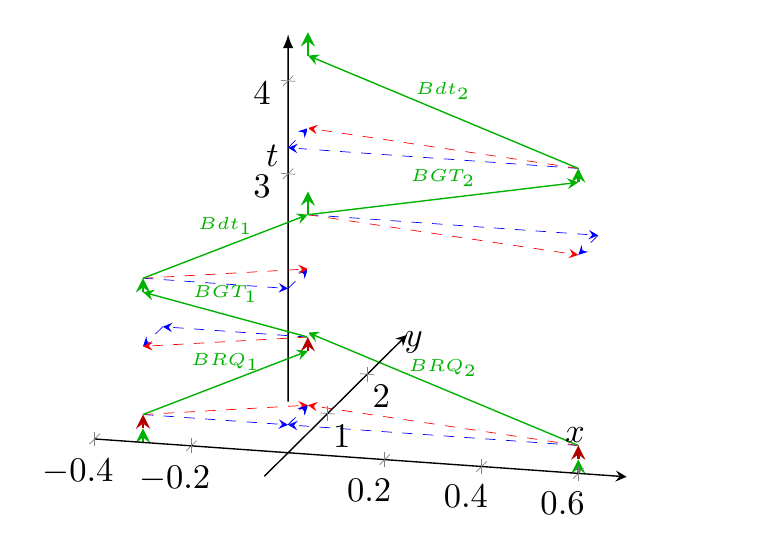}
		\caption{The operation of the sequential bus, in time-space coordinate system system. Near to axis \textit{t}, \textit{the lack of vertical arrows} signals "idle waiting" time\label{fig:Relativisticbus}}
	\end{figure}
	
	Under a biology-mimicking workload, 
	the second neuron must wait for all its inputs originating in the hidden layer.
	If we have $L$ neurons in the hidden layer,
	the transmission time of the neuron behind the hidden layer is $T_t=L\cdot 2\cdot T_B + T_d +X$. 
	This temporal behavior explains why "\textit{shallow networks with many neurons per layer \dots scale worse than deep networks with less neurons}"\cite{DeepNeuralNetworkTraining:2016}:
	\textit{the physical bus delivery time $T_d$, as well as the processing time $T_p$,
		become marginal if the layer forces to make
		many arbitrations to reach the bus}: the number of the 
	neurons in the hidden layer defines the transfer time   (Recall Figs.~\ref{fig:RelativisticComputation} and \ref{fig:FP16vsFP64HPCG} for the consequences of increasing the transfer time).
	In deeper networks, the system sends its messages at different times in its different layers
	(and, even they may have independent buses between the layers), although \textit{the shared bus persists in limiting the communication}.
	Notice that there is no way to organize the message traffic: only one bus exists.
	
	Fig.~\ref{fig:Relativisticbus} discusses, in terms of "temporal logic", the case depicted in the inset in Fig.~\ref{fig:Neuronal} (where the same operation is discussed in conventional terms): why using high-speed buses for connecting modern computer components leads to very severe performance loss, especially when one attempts to imitate neuromorphic operation. The two neurons of the hidden layer are positioned at (-0.3,0) and (0.6,0). The bus is at a position (0,0.5). The two neurons make their computation 
	(green arrows at the position of neurons), then they want to tell their result to their fellow neurons.
	Unlike in biology, first, they must have access to the shared bus (red arrows). The bus requests
	must reach the arbiter, that needs time, and so does the grant signal. 
	The core at (-.3,0) is closer to the bus, so its request is granted.
	As soon as the grant signal reaches the requesting core, the bus operation is initiated, and the data starts to travel to the bus. As soon as it reaches the bus, it is forwarded by the high speed of the bus. At that point, the other core's bus request is granted, and finally, the computed result of the second neuron is bused. 
	
	At this point comes into picture the role of the workload on the system: the two neurons in the hidden layer want to use the single shared bus, at the same time, for communication. As a consequence, 
	\textit{the apparent processing time is several times higher, than the physical processing time,
	and it increases linearly with the number of neurons in the hidden layer} (and, maybe,	with also the total number of neurons in the system, if a single high-speed bus is used).

The ratio of the time spent with forwarding data on the high-speed bus gradually decreases as the system's size increases. 
\textit{In vast systems, especially when attempting to mimic neuromorphic workload, 
	the speed of the bus is getting marginal}.
Notice that the times shown in the figure are not proportional: 
the (temporal) distances between cores are in the several picoseconds range,
while the bus (and the arbiter) are at a distance well above nanoseconds, so \textit{the actual temporal
	behavior (and the idle time stemming from it) is much worse than the figure suggests}. \textit{"The idea of using the popular shared bus to implement the communication medium is no longer acceptable, mainly due to its high contention."}~\cite{ReconfigurableAdaptive2016}.
The extraordinary workload of AI, makes it much harder to operate the systems.

When imitating biological processes, one needs to consider both the time at which the event can be "seen" in wet neurobiology (the biological time) and the time duration that the computer processor needs to deliver the result corresponding to the biological event (computing time)\footnote{The "wall-clock" time, reflecting computer operational details, is not considered here}.  	
The computing objects, intending to imitate biological systems, need to be aware of both time scales.
As discussed above, in connection with the serial bus, 
the technical implementation may introduce enormously 
low payload computing efficiency, and 
considerably distorts the time relations between computing and data delivery times.

Passing of time is measured via counting some periodic events,
such as clock periods in computing systems or spiking events in biological systems. In this event-based world, everything happening between those events happens "at the same time". However, the technical implementation (including measuring biological processes) may introduce another, unintended, granularity. The biological neurons perform analog integration; the technological implementations are prepared to perform
"step-wise" digital integration. This step involves (mostly) losing phase information. Furthermore, as detailed in~\textbf{\cite{VeghBrainAmdahl:2019}}, it introduces severe payload performance limits for neuronal operations.

\subsection{Training ANNs}

During training, we start showing an input, and the system begins to work, using the initial values of its synaptic weights. Those weights may be randomized, may be set according to some presumption, or maybe that they correspond to the previous input data. The signals that
the system sends are correct, but a receiver does not know the future: a signal must be physically delivered\footnote{Even if the message envelope contains a time stamp} before it can be processed.
Before that time, the neurons (and their dependents) may start to adjust their weights to a still undefined state: there is no synchronization.

Spiking is also a "look at me" signal in biology (unlike in the case of "technical neurons"): the computed feedback shall be sent to \textit{that} neuron,
reflecting the change its output caused. Without this, neurons receive feedback about "the
effect of all fellow neurons, including me"\footnote{Maybe it is worth to re-discuss, whether in "spatiotemporal" systems full or partial derivates shall be used.}. Receiving a spike, defines the beginning of the time of
validity of the signal; "leaking" also defines its "expiration time". When using spiking networks, their temporal behavior is vital.

In the example in~\textbf{\cite{VeghTemporal:2020}}, in the one-bit adder, the first AND gate has quite a short
indefinite time, but the OR has a long one. Neuronal operations show a similar behavior concerning having undefined states and weights, 
although they are more complex than a simple adder, and their operations are much longer.
Essentially, their operation is an iteration, where \textit{the actors mostly use mostly undefined input signals, 	and surely adjust their weights to false signals initially, and with a significant time delay at later times}. 
\textit{Not considering temporal behavior leads to painfully slow and doubtful convergence.} The larger is the system, the slower is its convergence, and the higher is the chance of "over-fitting".

Computing neuronal feedback results in a faster way, 
cannot help much, if at all. 
Delivering feedback information also needs time
and uses the same shared medium,
with all its disadvantages. In biology, the "computing time" and "communication time" are in the same
order of magnitude. In its technical implementation, the communication time is very much longer than computation. That is,
\textit{the received feedback refers to a time} (and related state variables) \textit{that was valid a very long time ago}.

In excessive systems, to provide seemingly higher performance,
some result/ feedback events must be dropped because of their long queueing.
Given that the feedback is computed from the results of the neuron that receives feedback,  the physical implementation of the computing system converts the logical dependence to time dependence~\textbf{\cite{VeghTemporal:2020}}.
Because of this time sequence, feedback messages will arrive to the neuron at a later \textit{physical time} (even if at the same \textit{biological time}, according to
their time stamp they carry),
so they stand at the end of the input message queue.
Because of this, it is highly probable that they "\textit{are dropped if the receiving process is busy over several delivery cycles}"~\cite{NeuralNetworkPerformance:2018}. 
\textit{In vast systems, feedback in the learning process involves results based on undefined inputs,
	furthermore, the calculated (and maybe correct) feedback may be neglected}.

The statements above are underpinned in other experimental investigations, too.
Introducing the spatio-temporal behavior of ANNs, even in its simple form, using separated (i.e., not connected in the way proposed here)  time and space contributions to describe them, "factorizing the 3D convolutional filters into separate spatial and temporal components yields significant gains in accuracy" and efficiency of
video analysis~\cite{SpatiotemporalLearning:2018,SpatiotemporalAction:2018}.
Furthermore, another careful analysis also discovered, that
"\textit{Yet the task of training such networks remains a challenging optimization problem. Several related problems arise: very long training time (several weeks on modern computers, for some problems), the potential for over-fitting (whereby the learned function is too specific to the training data and generalizes poorly to unseen data), and more technically, the vanishing gradient problem}". 
It is correctly stated that one of the reasons of the issues is the communication load, so it is a plausible attempt to 
reduce the number of communication units (BTW: the idea mimics the way as biology works):
"\textit{The immediate effect of activating fewer units is that propagating information through the network will be faster, both at training and at test time.}"\cite{ConditionalComputationBengion:2016}
However, confusing the biological and the computational times distorts their timing relations.
This also means that the computed feedback, based maybe on undefined inputs, reaches the previous layer's neurons faster.
A natural consequence is that (see their Fig.~5): "\textit{As $\lambda_s$ increases, the running time decreases, but so does performance.}"
The role of time (mismatching) is confirmed directly,
via making investigations in the time domain. "\textit{The CNN models are more sensitive to low-frequency channels than high-frequency channels}"~\cite{LearningFrequencyDomain:2020}:
the feedback can follow the changes in function of the
speed of the changes compared to the speed of feedback calculation.
	
\section{Conclusions}
To understand a neural network~\cite{UnderstandNeuralNetwork:2019} means not only correct coding and using proper weights:
considering the timing relations properly are at least as crucial. The larger the systems, the more crucial. The timely behavior is the key to learning and development. Both describing them mathematically  and implementing them technically, without considering their true time dependence, they mimic something different. That approaches result in "\textit{that any
	studies on processes like plasticity, learning, and development
	exhibited over hours and days of biological time are outside our
	reach}".~\cite{NeuralNetworkPerformance:2018}

The temporal behavior is a crucial attribute of neuronal operation, both in biological and technical computing systems.
The technical implementations lack synchronization, so they confuse the biological and computational time scales.
When imitating biological computing operation in technical computing systems, the time relations may be drastically distorted, which leads to unrealistic imitation of the biological behavior.
In the first round, 
the proposed mathematical handling enables us to find the reasons for the inefficient/erroneous/slow operation of the artificial neuronal systems. The second round helps to prepare systems with much higher efficacy and (from a biological point of view) correct operation.

\section*{Acknowledgements}
The authors thank Prof. P\'eter Somogyi for valuable comments on a previous version of the manuscript. 
Project no. 136496  has been implemented with the support provided from the National Research, Development and Innovation Fund of Hungary, financed under the K funding scheme. 

\section*{Conflict of interest statement}

The authors declare that no competing interests exist.

\section*{References}


\end{document}